\PassOptionsToPackage{unicode}{hyperref}
\PassOptionsToPackage{hyphens}{url}
\PassOptionsToPackage{dvipsnames,svgnames,x11names}{xcolor}
\documentclass[
  12pt]{article}

\usepackage{amsmath,amssymb}
\usepackage{iftex}
\ifPDFTeX
  \usepackage[T1]{fontenc}
  \usepackage[utf8]{inputenc}
  \usepackage{textcomp} % provide euro and other symbols
\else % if luatex or xetex
  \usepackage{unicode-math}
  \defaultfontfeatures{Scale=MatchLowercase}
  \defaultfontfeatures[\rmfamily]{Ligatures=TeX,Scale=1}
\fi
\usepackage{lmodern}
\ifPDFTeX\else
\fi
\IfFileExists{upquote.sty}{\usepackage{upquote}}{}
\IfFileExists{microtype.sty}{% use microtype if available
  \usepackage[]{microtype}
  \UseMicrotypeSet[protrusion]{basicmath} % disable protrusion for tt fonts
}{}
\makeatletter
\@ifundefined{KOMAClassName}{% if non-KOMA class
  \IfFileExists{parskip.sty}{%
    \usepackage{parskip}
  }{% else
    \setlength{\parindent}{0pt}
    \setlength{\parskip}{6pt plus 2pt minus 1pt}}
}{% if KOMA class
  \KOMAoptions{parskip=half}}
\makeatother
\usepackage{xcolor}
\makeatletter
\ifx\paragraph\undefined\else
  \let\oldparagraph\paragraph
  \renewcommand{\paragraph}{
    \@ifstar
      \xxxParagraphStar
      \xxxParagraphNoStar
  }
  \newcommand{\xxxParagraphStar}[1]{\oldparagraph*{#1}\mbox{}}
  \newcommand{\xxxParagraphNoStar}[1]{\oldparagraph{#1}\mbox{}}
\fi
\ifx\subparagraph\undefined\else
  \let\oldsubparagraph\subparagraph
  \renewcommand{\subparagraph}{
    \@ifstar
      \xxxSubParagraphStar
      \xxxSubParagraphNoStar
  }
  \newcommand{\xxxSubParagraphStar}[1]{\oldsubparagraph*{#1}\mbox{}}
  \newcommand{\xxxSubParagraphNoStar}[1]{\oldsubparagraph{#1}\mbox{}}
\fi
\makeatother

\usepackage{longtable,booktabs,array}
\usepackage{calc} % for calculating minipage widths
\usepackage{etoolbox}
\makeatletter
\patchcmd\longtable{\par}{\if@noskipsec\mbox{}\fi\par}{}{}
\makeatother
\IfFileExists{footnotehyper.sty}{\usepackage{footnotehyper}}{\usepackage{footnote}}
\makesavenoteenv{longtable}
\usepackage{graphicx}
\makeatletter
\def\maxwidth{\ifdim\Gin@nat@width>\linewidth\linewidth\else\Gin@nat@width\fi}
\def\maxheight{\ifdim\Gin@nat@height>\textheight\textheight\else\Gin@nat@height\fi}
\makeatother
\setkeys{Gin}{width=\maxwidth,height=\maxheight,keepaspectratio}
\makeatletter
\def\fps@figure{htbp}
\makeatother

\makeatletter
\@ifpackageloaded{caption}{}{\usepackage{caption}}
\AtBeginDocument{%
\ifdefined\contentsname
  \renewcommand*\contentsname{Table of contents}
\else
  \newcommand\contentsname{Table of contents}
\fi
\ifdefined\listfigurename
  \renewcommand*\listfigurename{List of Figures}
\else
  \newcommand\listfigurename{List of Figures}
\fi
\ifdefined\listtablename
  \renewcommand*\listtablename{List of Tables}
\else
  \newcommand\listtablename{List of Tables}
\fi
\ifdefined\figurename
  \renewcommand*\figurename{Figure}
\else
  \newcommand\figurename{Figure}
\fi
\ifdefined\tablename
  \renewcommand*\tablename{Table}
\else
  \newcommand\tablename{Table}
\fi
}
\@ifpackageloaded{float}{}{\usepackage{float}}
\floatstyle{ruled}
\@ifundefined{c@chapter}{\newfloat{codelisting}{h}{lop}}{\newfloat{codelisting}{h}{lop}[chapter]}
\floatname{codelisting}{Listing}

\makeatother
\makeatletter
\@ifpackageloaded{caption}{}{\usepackage{caption}}
\@ifpackageloaded{subcaption}{}{\usepackage{subcaption}}
\makeatother

\ifLuaTeX
  \usepackage{selnolig}  % disable illegal ligatures
\fi
\usepackage[]{natbib}
\usepackage{bookmark}

\IfFileExists{xurl.sty}{\usepackage{xurl}}{} % add URL line breaks if available
\hypersetup{
  pdftitle={Title},
  pdfauthor={Author 1; Author 2},
  pdfkeywords={3 to 6 keywords, that do not appear in the title},
  colorlinks=true,
  linkcolor={blue},
  filecolor={Maroon},
  citecolor={Blue},
  urlcolor={Blue},
  pdfcreator={LaTeX via pandoc}}

\newcommand{\anon}{1}

\newtheorem{theorem}{Theorem}
\newtheorem{lemma}{Lemma}
\newtheorem{proposition}[theorem]{Proposition}

\makeatletter
\def\thanks#1{\protected@xdef\@thanks{\@thanks
        \protect\footnotetext{#1}}}
\makeatother

\usepackage{tikz}
\usetikzlibrary{arrows,positioning,shapes,calc}

\tikzset{
    node/.style={rectangle, draw, fill=white, minimum size=1.5cm, text centered},
    treatment/.style={node, fill=blue!10}, % ´¦Àí±äÁ¿£¨Ç³À¶£©
    mark/.style={node, fill=red!10},       % ±ê¼Ç±äÁ¿£¨Ç³ºì£©
    outcome/.style={node, fill=green!10},  % ½á¹û±äÁ¿£¨Ç³ÂÌ£©
    intervention/.style={node, fill=yellow!10, diamond}, % ¸ÉÔ¤£¨Ç³»ÆÁâÐΣ©
    arrow/.style={->, thick, >=stealth}
}

\DeclareMathOperator{\E}{\mathbb{E}}
\begin{document}

\def\spacingset#1{\renewcommand{\baselinestretch}%
{#1}\small\normalsize}

%%%%%%%%%%%%%%%%%%%%%%%%%%%%%%%%%%%%%%%%%%%%%%%%%%%%%%%%%%%%%%%%%%%%%%%%%%%%%%

\if1\anon
{
  \title{\bf Causal inference for censored data with continuous marks}
  \author{Lianqiang Qu,~~Long Lv\\
  %\thanks{Lianqiang Qu's research was partly supported by the National Natural Science Foundation of China (No. 12471256).}\\
    School of Mathematics and Statistics, Central China Normal University\\
    Wuhan, Hubei 430079, China\\
    Liuquan Sun\\
    %\thanks{Liuquan Sun is the corresponding author. Liuquan Sun's research was partly supported by the National Natural Science Foundation of China (No. 12571299).}\\
    SKLMS, Academy of Mathematics and Systems Science, \\
    Chinese Academy of Sciences,\\
and School of Mathematical Sciences, University of Chinese Academy of Sciences,\\
Beijing, 100190, China}
  \maketitle
} \fi

\if0\anon
{
  \bigskip
  \bigskip
  \bigskip
  \begin{center}
    {\LARGE\bf Causal inference for censored data with continuous marks}
\end{center}
  \medskip
} \fi

\bigskip

\spacingset{1.5}

\begin{abstract}
This paper presents a framework for causal inference in the presence of censored data,
where the failure time is marked by a continuous variable referred to as a mark.
The mark is observed after treatment and is not meaningful when the failure time is censored.
In addition, due to the continuous nature of the marks, observations at each given mark are sparse.
These facts make the identification and estimation of causality a challenging task.
To address these issues, we define a new mark-specific treatment effect within the potential outcomes framework
and characterize its identifying conditions.
We then propose a local smoothing estimator for the causal effects
and establish its asymptotic properties.
We further develop testing methods to evaluate whether the treatment has an effect
on the failure time when controlling the values of the mark at certain points or within a defined interval,
and develop a Gaussian approximation method to obtain the critical values.
We evaluate our method using simulation studies as well as a real dataset from the Antibody Mediated Prevention trials.
\end{abstract}

\noindent%
{\it Keywords:} Causal effect; Potential outcomes; Post-treatment variable; Survival data.
\vfill

\spacingset{1.8} % DON'T change the spacing!

\section{Introduction}\label{sec-intro}

Causal inference has played a crucial role in many domains, such as statistics, computer science,
education, public policy, and economics, for decades.
Unlike correlation, it seeks to draw a conclusion about a causal
connection based on the conditions of the occurrence of a treatment--such as
whether a policy genuinely improves outcomes or if a drug leads to recovery.
Therefore, a thorough understanding of causality is essential for effective decision-making.

\subsection{Motivation and contributions}

In various applications, the event time of interest, commonly termed ``failure time",
is often subject to censoring.
In addition, the failure time is usually marked by a continuous random variable, known as a mark.
For example, subjects in Antibody Mediated Prevention (AMP) trials were randomly assigned to receive
either infusions of the monoclonal broadly neutralizing antibody (bnAb) VRC01 or a placebo.
The mark variable, 80\% inhibitory concentration (IC80),
measures the HIV-1 Envelope neutralization sensitivity to VRC01.
As IC80 is unique for each infected subject
and the genetic diversity of HIV-1 is considerable,
the mark is a continuous random variable.
A natural and compelling question is:

{\it Does a treatment (e.g., VRC01) have a causal effect
on the failure time (e.g., time to HIV infection) marked by a continuous random variable (e.g., IC80)?}

\noindent To address this question, we need to
(1) define the causal effect of a treatment when outcomes are marked by a continuous variable;
and (2) develop estimation and inference methods for such causal effects.

In our context, the challenges related to the definition, identification and estimation of causality are as follows.
First, the mark is observed after treatment, which is termed a post-treatment variable.
If we directly treat it as a baseline covariate,
it may introduce selection bias and hinder a clear causal interpretation.
Second, there is an observational dependency between the failure times and the marks.
Specifically, during the observation period, subjects are potentially at risk of exposure to various viral strains.
The mark is recorded only upon the occurrence of HIV-1 infection;
in cases where infection does not occur, the mark is undefined and lacks meaningful interpretation.
Moreover, the prevention efficacy of bnAb (vs. placebo) against HIV-1 diagnosis varies with IC80s \citep{Corey2021, J2024},
indicating heterogeneous treatment effects.
These facts complicate the definition and identification of causality.
Furthermore, since the mark is continuous, observations at any given mark are extremely sparse.
Direct estimation will produce a significant bias for the causal effect.
% which may lead to observations at a given mark being present in the treatment group but absent in the control group.
%which may lead to a lack of comparable benchmarks for estimating the causal effect at these marks.
%Sparse observations can also lead to highly biased estimates for causality.
%due to the identifiability issues associated with the mark under censoring.
%We note that the role of the mark is analogous to that of the failure cause
%within the competing risks framework \citep{Prentice1978}.
%While causal inference methods have been studied for competing risks data
%with discrete failure causes (e.g., \citealp{Y2020,SY2022,RL2024}),
%existing methods require dense observations at a given cause.
%This condition differs significantly from ours,
%rendering existing methods inapplicable to our context.
%We note that the role of the mark is analogous to that of the failure cause
%within the competing risks framework \citep{Prentice1978}.
%While causal inference methods have been studied for competing risks data
%with discrete failure causes (e.g., \citealp{Y2020,SH2021,SY2022,RL2024}),
%existing methods require dense observations at a given cause.
%This condition significantly differs from ours,
%rendering existing methods inapplicable to our context.

To address these issues, we characterize a mark-specific treatment effect and its identifying conditions
within the potential outcomes framework (\citealp{Neyman1923,R1974}).
Let $A\in\{0,1\}$ denote a treatment indicator:
1 for the treatment and 0 for the control.
As the mark is a post-treatment variable and can be treated as a cause of the failure \citep{SGM2009,JG2013,QSS2024},
we define $V(a)$ as the potential mark if a subject receives treatment $A=a$
and define $T(a,v)$ as the potential failure time of interest
if a subject receives treatment $A=a$ and the potential mark is specified at $V(a)=v$.
We assume that the support of $V(a)$ is $[0,1]$ for $a\in\{0, 1\}$.
Following our analysis of motivation data,
we consider a randomness experiment with an assignment mechanism defined by $\mathbb{P}(A=1)=\pi$.
%Define $V=V(A)$, $T=T(A,V)$ and $Y=\min\{T, C\}$.
%Further define $\Delta=I\{T\le C\}$,
%where $I(\cdot)$ denotes an indictor function.
%%Let $V$ denote the mark, which is observed only when $\Delta=1$.
%Note that $V$ is observed only when $\Delta=1$.
%then $V$ is not meaningful.
%where $Y_i=\min\{T_i, C_i\}$ with $T_i=T_i(A_i)$ and $\Delta_i=I\{T_i\le C_i\}$.
%where $\pi\in(0,1)$ denotes the treatment probability.
%For a comprehensive review of completely randomized experiments in causal inference,
%we refer to \cite{IR2015} and \cite{Ding2024}.

Let $[b_L,b_U]\subset[0,1]$ be the subinterval of interest.
For each given $v\in[b_L,b_U]$, define
$$
\tau_a(v)=\mathbb{E}[T(a,v)],
$$
where $\tau_a(v)$ denotes the average potential failure time if $A$ is set to $a$
and the mark to the value $v$.
In AMP trials, $\tau_a(v)$ represents the average potential infection time
caused by HIV-1 strains marked by an IC80 level of $v$ within the group $a$.
In addition, $\tau_a(v)$ can vary with $v$, capturing heterogeneous treatment effects.
%Note that $\tau_a(v)$ differs from classical nonparametric regression models,
%Specifically, $\tau_a(v)$ is defined based on the joint distribution of the potential failure time and the mark,
%which focus on modeling the conditional distribution of the outcome given baseline covariates.
Define
\begin{align*}
\tau(v)=\tau_1(v)-\tau_0(v),
\end{align*}
which captures the mean difference of the potential failure times
when the mark is set to $v$ uniformly in the population.
Based on $\tau(v)$, we define the following mark-specific treatment effects:
%\begin{definition}[Mark-specific causal effects]~
\begin{itemize}
\item Local treatment effect: If $\tau(v)\neq 0$ for any fixed $v\in [b_L,b_U]$,
we say that $A$ has a local treatment effect with the mark set to $v$.

\item Global treatment effect: If $\tau(v)\not\equiv 0$ over an interval $v\in[b_L,b_U]$,
we say that $A$ has a global treatment effect over $v\in[b_L,b_U]$.

\item Heterogeneous treatment effect: If $\tau(v)\not\equiv \psi$ in an interval $v\in[b_L,b_U]$,
where $\psi$ denotes a constant, we say that $A$ has a heterogeneous treatment effect with respect to $v$ over $[b_L,b_U]$.
\end{itemize}

In this paper, we attempt to address the aforementioned questions
by analyzing the identifying conditions of the mark-specific treatment effects and developing methods
for their estimation and inference.
The main contributions are summarized as follows.

First, we propose a framework for analyzing causal inference
in which the failure time is marked by a continuous variable.
This framework incorporates the mark into the definition of causal effects
and accounts for the potential heterogeneity in the treatment effects induced by the mark.
Under regular assumptions, we address the issue of identifiability of the mark-specific treatment effects.

Second, to address the issue of sparse observations, we develop a local smoothing method to estimate $\tau(v)$,
which combines kernel smoothing techniques with inverse probability weighting methods.
The proposed method leverages information from data within a neighborhood of the marks,
effectively capturing the local similarities in effects of the treatment.
To enhance the practicality of our method,
we also develop a data-driven procedure to estimate the bandwidth.

Third, we establish the uniform consistency and asymptotic normality of the proposed estimator.
In addition, we propose a consistent estimator for the asymptotic variance of the estimators,
providing a reliable basis for inference.
Moreover, we develop testing methods to assess whether treatment exhibits local treatment effects,
global treatment effects, and heterogeneous treatment effects.
We approximate the null distributions of the proposed tests by a sequence of suprema of Gaussian processes.
Based on this Gaussian approximation, we develop a Monte Carlo method to obtain the critical values.
% drive the upper bound for this approximation.

\subsection{Related works}

The current research is closely related to two important areas:
causal inference with censored data and causal mediation analysis.
Various methods have been developed to facilitate causal inference with censored data
(e.g., \citealp{CS2007,TT2014,TA2014,Yu2015,WY2021,SY2022,LS2023,Cui2023,MV2025}).
However, the definition of causal effect in existing methods overlooks the impact of the mark.
When the mark is present, but not taken into account,
it may fail to demonstrate significant causality.
A toy example to illustrate this issue is provided in Section \ref{sec3}.
\cite{SGM2009} investigated the proportional hazards model with a continuous mark,
while \cite{QSS2024} proposed a mark-specific quantile regression model to analyze such data.
Additional related works include \cite{SLG2013}, \cite{JG2013, JG2016}, \cite{HSS2017}, and \cite{SQHG2020}, among others.
However, these methods excel at estimating associations between a treatment and the failure time
but fall short of establishing causal implications.

In the causal mediation literature, $\tau(v)$ is referred to as the controlled direct effect \citep{RG1992,P2001},
in which the potential mark corresponds to a potential mediator.
In this study, we do not focus on decomposing the total effect into indirect and direct effects \citep{VV2009, IKT2010, Y2020, Q2024, Z2025}.
Instead, our goal is to provide a testing method to evaluate
whether the treatment has an effect on the failure times
while controlling its corresponding mark values at certain points or within a defined interval.
For example, VRC01 may not demonstrate a more effective prevention of overall HIV-1 acquisition compared to a placebo;
however, it can prevent infections caused by VRC01-sensitive HIV-1 (IC80 $\le 1\mu\mathrm{g}$ per milliliter) \citep{Corey2021}.
In addition, in certain cases, it may fail to identify the treatment effect without controlling the mark,
even when considering the decomposition of the total effect into indirect and direct effects;
see Section \ref{sec3} for a toy example.
Therefore, in the current work, we focus on the identification and estimation of the controlled direct effect.
Moreover, the potential mediator is assumed to be fully observed in the existing causal mediator analysis,
and there is no observational dependency between the outcomes and the mediator,
particularly when the mediator is a continuous variable.
Therefore, we require the development of a new method for causal inference in this context.

\subsection{Organization of the article}
The remainder of this paper is organized as follows.
Section \ref{IE} presents regular conditions for identifying $\tau(v)$
and introduces a local smoothing estimator for $\tau(v)$.
We establish the consistency and asymptotic normality of the estimators in Section \ref{AsyP},
while presenting a data-driven bandwidth selection method in Section \ref{BS}.
Section \ref{sec3} presents testing methods for the causal effects.
Section \ref{Section:SS} conducts simulation studies to evaluate the finite-sample performance of the proposed methods,
while Section \ref{realdata} illustrates these methods using a data set from the AMP trials.
Discussions are provided in Section \ref{sec-discussion}.
The proofs of our main results are presented in the Appendix.

\section{Main results}
\subsection{Identification and estimation}
\label{IE}
In this section, we first consider the identifiability issue of $\tau(v)$
and then develop a kernel smoothing method to estimate it.
Let $C$ be the censoring time.
Define $V=V(A)$ and $T=T(A, V(A))$.
Let $Y=\min\{T,C\}$ be the observed time and $\Delta=I(T\le C)$ be the censored indicator,
where $I(\cdot)$ denotes the indicator function.
Note that $V$ is observed only when $\Delta=1$.
Assume that we observe $n$ independent and identically distributed samples of $(Y,\Delta, \Delta V,A)$,
denoted as $\{(Y_i,\Delta_i,\Delta_iV_i,A_i): 1\le i\le n\}$.
Let $S_a(t)=\mathbb{P}(C\ge t|A=a)$ denote the survival function of $C$
and $f_a(v)$ denote the conditional distribution of $V$ conditional on $A=a$.

For the identification of $\tau(v)$, we consider the following conditions.

%\begin{condition}[Identification]~For $a\in\{0,1\}$, assume that
%\begin{itemize}
%\item[](i) (Independent censoring). $C(a)$ is independent of $(T(a), V(a))$.
%\item[](ii) (Positivity). $0<P(A=1)=\pi<1$ and $S_a(T(a))>0$ almost surely.
%\item[](iii) (Unconfoundedness). $A$ is independent of $(T(a), V(a), C(a))$.
%\item[](iv) (Stable Subject Treatment Value Assumption). The potential outcomes of subject $i$ do not depend on other subjects' treatments.
%\item[](v) (Consistency). $(T,V,C)=(T(a),V(a),C(a))$ almost surely.
%\end{itemize}\label{Id}
%\end{condition}

\noindent
{\bf Condition 1} (Identification). For $a\in\{0,1\}$, assume that\\
(i) Ignorability: $A$ is independent of $T(a,v)$ and $V(a)$ is conditionally independent of $T(a,v)$ given $A=a$.\\
(ii) No-interference: The potential outcomes of a subject do not depend on other subjects' treatments.\\
(iii) Consistency: $V=V(a)$ almost surely if $A=a$. In addition, $T=T(a,v)$ almost surely if $A=a$ and $V(a)=v$.\\
(iv) Positivity: $0<\pi<1$, $f_a(v)>f_{\min}>0$, and $S_a(T)>s_{\min}>0$ almost surely.\\
(v) Independent censoring: $C$ is independent of $T$ and $V$ conditionally on $A$.

%Condition 1 is widely adopted in the potential outcome framework.
The ignorability assumption implies that the potential outcome under levels $a$ and $V(a)=v$ is the same across treatment levels.
The no-interference implies that treating one subject does not influence the outcomes for another subject in the study population
(e.g., due to spillover or peer effects).
The consistency says that there are no other versions of the treatment
and the outcomes are clearly defined.
The consistency and the no-interference ensure
that potential outcomes are uniquely defined by a subject's own treatment.
The positivity assumption means that treatment is not assigned deterministically,
in the sense that every subject has a chance of receiving treatment level $a$
and a chance that their failure time is caused at the make level $v$.
The independent censoring assumption is widely adopted in survival data analysis.
%this can be a particularly strong assumption with continuous treatments.

Under Condition 1, we have
\begin{align}\label{identification-tau1}
\tau_a(v)=&\mathbb{E}[T(a,v)|A=a,V(a)=v]~~\quad\qquad\qquad\qquad(\text{by~Condition~1 (i)})\nonumber\\
=&\mathbb{E}[T|A=a,V=v]~\quad\qquad\qquad\qquad\qquad\qquad(\text{by~Conditions~1 (ii)~and~1 (iii)})\nonumber\\
%%=&\mathbb{E}[T|A=a,V=v]~~\quad\qquad\qquad(\text{by~ Conditions~1 (ii)~and~1 (iii)})\nonumber\\
=&\lim_{h\rightarrow 0}\frac{\mathbb{E}[TI(|V-v|<h)|A=a]}{\mathbb{E}[I(|V-v|<h)|A=a]}\qquad\qquad\qquad
(\text{by~Condition 1 (iv)})\nonumber\\
=&\lim_{h\rightarrow 0}\frac{\mathbb{E}[\Delta YI(|\Delta V-v|<h)/S_a(Y)|A=a]}
{\mathbb{E}[\Delta I(|\Delta V-v|<h)/S_a(Y)|A=a]}\quad(\text{by~Conditions~1 (iv)~and~1 (v)}).
\end{align}
The third equation in \eqref{identification-tau1} implies that under Condition 1,
we can capture the observational dependence between the failure time $T$ and the mark $V$,
while the last equation addresses the issue of censoring for both $T$ and $V$.
By \eqref{identification-tau1}, $\tau_a(v)$ can be identified using the observations $\{(Y_i,\Delta_i,\Delta_iV_i,A_i): 1\le i\le n\}$.

Now, we consider the estimation of $\tau_a(v)$.
Due to the sparsity of observations at any given mark,
directly estimating $\tau_a(v)$ is impractical.
To address this issue, we assume that $\tau_a(v)$ is continuous,
such that $\tau_a(u)\approx \tau_a(v)$ when $u$ is close to $v$.
This assumption can be interpreted as follows:
the time to infection is highly similar
when the infection is caused by HIV-1 strains with comparable IC80s.
Under this assumption, we can estimate $\tau_a(v)$ by leveraging information from observations with marks near $v$.
Specifically, let $L$ be the follow-up time, $n_a$ denote the number of subjects in the group $a$
and $N_i(t,u)=\Delta_iI(Y_i\le t, \Delta_iV_i\le u)$.
By the last equation in \eqref{identification-tau1}, for each $v\in[b_L,b_U]$ and a small $h$, we can estimate $\tau_a(v)$ using
%$$
%\widehat\tau_a(v)=\frac{\widehat{\mathbb{E}}[\Delta YI(|\Delta V-v|<h)/S_a(Y)|A=a]}{\widehat{\mathbb{E}}[\Delta I(|\Delta V-v|<h)/S_a(Y)|A=a]},
%$$
\begin{align*}
\widehat\tau_a(v)=&\frac{\sum_{i: A_i=a}\Delta_iY_iK_h(\Delta_iV_i-v)/\widehat S_a(Y_i)}
{\sum_{i: A_i=a}\Delta_iK_h(\Delta_iV_i-v)/\widehat S_a(Y_i)}\\
=&\frac{\sum_{i: A_i=a}\int_{0}^1\int_{0}^{L}\widehat S_a^{-1}(t)tK_h(u-v)N_i(dt,du)}
{\sum_{i: A_i=a}\int_{0}^1\int_{0}^{L}\widehat S_a^{-1}(t)K_h(u-v)N_i(dt,du)},
\end{align*}
%where
%\begin{align*}
%\widehat{\mathbb{E}}\bigg[\frac{\Delta}{S_a(Y)} YI(|\Delta V-v|<h)\bigg|A=a\bigg]
%&=\frac{1}{n_a}\sum_{i: A_i=a}\frac{\Delta_i}{\widehat S_a(Y_i)}Y_iK_h(\Delta_iV_i-v)\\
%&=\frac{1}{n_a}\sum_{i: A_i=a}\int_{0}^1\int_{0}^{L}\frac{t}{\widehat S_a(t)}K_h(u-v)N_i(dt,du),\\
%\widehat{\mathbb{E}}\bigg[\frac{\Delta}{S_a(Y)} I(|\Delta V-v|<h)\bigg|A=a\bigg]
%&=\frac{1}{n_a}\sum_{i: A_i=a}\frac{\Delta_i}{\widehat S_a(Y_i)}K_h(\Delta_iV_i-v)\\
%&=\frac{1}{n_a}\sum_{i: A_i=a}\int_{0}^1\int_{0}^{L}\frac{1}{\widehat S_a(t)}K_h(u-v)N_i(dt,du).
%\end{align*}
%\begin{align*}
%\widehat{\mathbb{E}}\bigg[\frac{\Delta}{S_a(Y)} YI(|\Delta V-v|<h)\bigg|A=a\bigg]
%&=\frac{1}{n_a}\sum_{i: A_i=a}\int_{0}^1\int_{0}^{L}\frac{t}{\widehat S_a(t)}K_h(u-v)N_i(dt,du),\\
%\widehat{\mathbb{E}}\bigg[\frac{\Delta}{S_a(Y)} I(|\Delta V-v|<h)\bigg|A=a\bigg]
%&=\frac{1}{n_a}\sum_{i: A_i=a}\int_{0}^1\int_{0}^{L}\frac{1}{\widehat S_a(t)}K_h(u-v)N_i(dt,du).
%\end{align*}
where $K_h(x)=K(x/h)/h,$ $K(x)$ is a kernel function and $h$ denotes a bandwidth.
Additionally, $\widehat S_a(t)$ denotes the Kaplan-Meier estimator of $S_a(t)$ using observations in the group $A=a$.
The use of the same bandwidth $h$ in both $\widehat{\tau}_0(v)$ and $\widehat{\tau}_1(v)$ is intended for analytical simplicity,
and it is  straightforward to extend the analysis to accommodate different bandwidths.
We note that our proposed estimator of $\widehat\tau_a(v)$ can be expressed as the following least squires estimator:
$$
\widehat\tau_a(v)=\arg\min_{b}~~\frac{1}{n_a}\sum_{i: A_i=a}\frac{\Delta_{i}}{\widehat S_a(Y_i)}(Y_i-b)^2K_h(\Delta_iV_i-v),
$$
which plays a critical role in the development of a data-driven bandwidth selection method; see Section \ref{BS}.
Based on $\widehat\tau_a(v)$, we introduce an estimator for $\tau(v)$ as follows:
\begin{align*}
\widehat\tau(v)=\widehat\tau_1(v)-\widehat\tau_0(v).
\end{align*}

\subsection{Asymptotical properties}\label{AsyP}

In this section, we establish the asymptotical properties of $\widehat\tau(v)$,
including its uniform consistency and asymptotic normality.
Define $f_a(t|v)$ as the conditional density function of $T$ given $V=v$ and $A=a$
and $f_{a}(t,v)=f_a(t|v)f_a(v)$.
We assume the following conditions.

\noindent {\bf Condition 2} (Smoothing).
The functions $\tau_a(v)$ and $f_a(t,v)$ have continuous second derivatives with respect to $v$ on $[0,1]$.
In addition, $f_a(t,v)$ is Lipschitz continuous.
That is, $|f_a(t_1,v_1)-f_a(t_2,v_2)|\le \eta_1|t_1-t_2|+\eta_2|v_1-v_2|$ for any $(t_1,v_1), (t_2,v_2)\in [0,L]\times[0,1]$.
Here, $\eta_1$ and $\eta_2$ are positive constants.

\noindent {\bf Condition 3} (Kernel and bandwidth).
The kernel $K(x)$ is a symmetric density function with support $[-1, 1],$
and has bounded variation.
Additionally, the bandwidth $h$ satisfies $h\rightarrow 0$ and $nh/\log n\rightarrow\infty$ as $n\rightarrow\infty$.

Condition 2 gives some restrictions on the smoothness of $\tau(v)$
and the conditional density $f_a(t,v)$ of $T$ and $V$ given $A=a$,
which are standard assumptions in the context of nonparametric methods.
The first part of Condition 3 restricts the selection of the kernel function.
This condition is satisfied by most common kernels,
including the uniform kernel $K(x)=I(|x|\le 1)/2$ and the Epanechnikov kernel $K(x)=(3/4)(1-x^2)I(|x|\le 1)$.
The second part assumes that the bandwidth $h$ should decrease with the sample size, but not too rapidly.
This is a standard requirement in kernel smoothing methods \citep{FG1996}.
Specifically, we require the bandwidth $h$ to approach zero to control the bias of $\widehat\tau(v)$,
while $nh$, which can be viewed as the local sample size, must tend to infinity to control the variance.
A more detailed discussion of the bandwidth $h$ is provided below,
along with the development of a data-driven method for its practical selection; see Section \ref{BS}.

The uniform consistency is given in the following theorem.

\begin{theorem}
\label{theo1}
Under Conditions 1-3, we have
$$
\sup_{v\in [b_L,b_U]}\|\widehat\tau(v)-\tau(v)\|=O_p\bigg(\sqrt{\frac{\log n}{nh}}+h^2+\frac{1}{\sqrt{n}}\bigg).
$$
% uniformly in $v\in [b_L,b_U].$
\end{theorem}
The proof of Theorem \ref{theo1} can be found in the Appendix.
Theorem \ref{theo1} indicates that the convergence rate consists of two components.
Specifically, $\sqrt{\log n/(nh)}+h^2$ represents the rate
achieved in standard non-parametric regression problems without the nuisance function $S_a(t)$.
If $h$ approaches zero very slowly,
$\sqrt{\log n/(nh)}$ decreases quickly.
In contrast, if $h$ approaches zero quickly,
then $h^2$ will also decrease quickly; however, $\sqrt{\log n/(nh)}$ will approach zero at a slower rate.
To balance the bias and variance, we require $h=O((n/\log n)^{-1/5})$.
This condition implies that both $\sqrt{\log n/(nh)}$ and $h^2$ are of the order $(n/\log n)^{-2/5}$,
which is the optimal convergence rate for standard non-parametric regression
with a single covariate when the regression function is twice continuously differentiable (e.g., \citealp{Tsybakov2009}, Theorem 2.10).
The second component, $n^{-1/2}$, is the convergence rate arising from the estimation of the nuisance function $S_a(t)$,
which exhibits a faster rate compared to the first component.

Next, we establish the asymptotic normality of $\widehat\tau(v)$.

\begin{theorem}
\label{theo2}
Under Conditions 1-3, for each $v\in[b_L,b_U],$
we have
$$
(nh)^{1/2}\{\widehat\tau(v)-\tau(v)-b_h(v)\}\stackrel{d}{\longrightarrow} N(0,\sigma^2(v)),
$$
where $b_h(v)=\nu_0\tau''(v)h^2/2$ and
\begin{align*}
\sigma^2(v)=\nu_1[\sigma_{1}^2(v)/\pi+\sigma_0^2(v)/(1-\pi)]
\end{align*}
with $\sigma_{a}^2(v)=f_a^{-1}(v)\int_0^{\infty}(t-\tau_a(v)^2f_a(t|v)S_a^{-1}(t)dt$ for $a\in\{0,1\}$.
Here, $\nu_0=\int_{-1}^1 u^2K(u)du$ and $\nu_1=\int_{-1}^1K^2(u)du$.
\end{theorem}

The proof of Theorem \ref{theo2} can be found in the Appendix.
Theorem \ref{theo2} indicates that the proposed estimator is asymptotically normal following appropriate scaling and centering.
However, the scaling is performed using the square root of the local sample size, $\sqrt{nh}$,
instead of the conventional parametric rate, $\sqrt{n}$.
Additionally, in standard non-parametric regression, the estimator may be consistent but not precisely centered at $\tau(v)$;
there exists a bias term $b_h(v)$ of the order $h^2$.
In the following, we assume $nh^5\rightarrow 0$ to ensure that $b_h(v)$ becomes asymptotically negligible,
which facilitates the construction of confidence intervals around $\tau(v)$.
Specifically, we can construct the confidence interval for $\tau(v)$ as follows:
\begin{align}\label{CI}
[\widehat\tau(v)-z_{\alpha/2}\widehat\sigma(v)/\sqrt{nh},\widehat\tau(v)+z_{\alpha/2}\widehat\sigma(v)/\sqrt{nh}],
\end{align}
where $z_{\alpha/2}$ denotes the $(1-\alpha/2)$-percentile of the standard normal distribution.
Here, $\widehat\sigma(v)$ denotes an estimator of $\sigma^2(v)$, defined as
$$
\widehat\sigma^2(v)=(nh)\sum_{a\in\{0,1\}}n_a^{-2}\sum_{i:A_i=a}\widehat\vartheta_{ai}^2(v),
$$
where
$$
\widehat\vartheta_{ai}(v)=\int_{0}^1\int_{0}^{L}\frac{t-\widehat\tau_a(v)}{\widehat f_a(v)\widehat S_a(t)}K_h(u-v)N_i(dt,du).
$$
Under Condition 1, using arguments similar to those in the proof of Theorem \ref{theo1},
we can demonstrate that $\widehat\sigma^2(v)$ is a uniformly consistent estimator for $\sigma^2(v)$.

\subsection{Bandwidth selection}\label{BS}
The selection of bandwidth $h$ is a critical part in estimating $\widehat\tau(v)$:
too much smoothing yields large biases and too little yields excessive variance.
In this subsection, we consider a data-driven method to estimate the bandwidth $h$.
We employ a unified selection approach from \cite{HHM1988},
which includes generalized cross-validation,
Akaike's information criterion, and leave-one-out cross-validation as special cases.
They showed the asymptotic equivalence and optimality of such approaches.
Since our estimator $\widehat\tau_a(v)$ can be obtained using the least squires method,
inspired by \cite{HHM1988}, we can estimate the bandwidth by
$$
\widehat h=\text{arg}\min_{h}~\sum_{a\in\{0,1\}}n_a^{-1}\sum_{i: A_i=a}\frac{\Delta_{i}}{\widehat S_a(Y_i)}\big(Y_i-\widehat\tau_a(V_i)\big)^2w(h)
$$
where $w(h)=1+2(nh)^{-1}K(0)$ is a correction factor.

An alternative method is cross-validation.
This approach requires to randomly divide the data into \( M \) parts.
For each bandwidth \( h \), we estimate the curve \( \tau_a(v) \) using \( M-1 \) training folds,
and then use these estimates to compute the prediction error on the test fold.
After repeating this process across all \( M \) folds,
we calculate the average of the prediction errors.
Finally, we determine the bandwidth that minimizes these errors.
However, this process may be computationally intensive.
%We can also use the rule of thumb bandwidth. Through Theorem \ref{theo2}, we choose
%$h=\varpi\hat\sigma_Vm^{-1/4}$,
%where $\hat\sigma_V$ is the estimated standard error of the observed marks,
%$m$ is the number of observed failure times, and
%$\varpi>0$ is a prespecified constant \citep{SGM2009}.

\section{Testing for causal effects}\label{sec3}
In this section, we consider the testing problems associated with $\tau(v)$.
%If the bnAb has no effects on the time to HIV-1 infection,
%then there is no significant difference between $\tau_1(v)$ and $\tau_0(v)$ for all $v \in[b_L, b_U]$:
%$\tau(v)\equiv 0$ for all $v \in[b_L, b_U].$
First, we consider testing the following hypothesis:
\begin{align*}
H_{0}^G: \tau(v)\equiv 0 \ \text{for}~v\in[b_L, b_U]\ \
\text{versus}\ \ H_{1}^G: \tau(v)\not\equiv0\ \text{for}~v\in[b_L, b_U].
\end{align*}
When $H_0^G$ is true, there is no difference between $\tau_1(v)$ and $\tau_0(v)$ over the interval $[b_L, b_U]$,
indicating that the effect of treatment on failure time is not significant.
To test $H_{0}^G,$ we consider the following statistic:
\begin{align*}
\mathcal{Z}=\sup_{v\in[b_L,b_U]} \sqrt{nh}|\widehat\tau(v)|/\widehat\sigma(v).
\end{align*}
The statistic $\mathcal{Z}$ approaches zero when $H_{0}^G$ is true.
Thus, we reject $H_{0}^G$ if $\mathcal{Z}>c_{1}(\alpha)$,
where $c_{1}(\alpha)$ denotes the critical value at the significance level $\alpha$.
To obtain the critical value $c_{1}(\alpha),$ we simulate the distribution of $\mathcal{Z}$
via a direct Monte Carlo method \citep{CCK2013, CCK2014}, which is given below.

%To obtain the critical value $\lambda_{1}(\alpha),$ we employ a resampling technique \citep{LWY1993}.
%By the proof of Theorem \ref{theo2}, under $H_{0}^G$, we have that $(nh)^{1/2}\widehat{\tau}(v)$
%is asymptotically equivalent to
%\begin{align}\label{Asym}
%\sqrt{\frac{h}{n}}\sum_{i=1}^n \bigg[\frac{A_i}{\widehat\pi}\widehat\vartheta_{1i}(v)-\frac{1-A_i}{1-\widehat\pi}\widehat\vartheta_{0i}(v)\bigg]
%\end{align}
%uniformly in $v\in[b_L,b_U]$.
%Define
%\begin{align*}
%\mathcal{G}^*=\max_{v\in[b_L,b_U]}\frac{h}{n}\bigg[\sum_{i=1}^n W_i\bigg\{\frac{A_i}{\widehat\pi}\widehat\vartheta_{1i}(v)-\frac{1-A_i}
%{1-\widehat\pi}\widehat\vartheta_{0i}(v)\bigg\}\bigg]^2\bigg/{\widehat\sigma^2(v)},
%\end{align*}
%where $W_i\ (i=1,\dots,n)$ are independent standard normal variables and are independent of the observed data.
%According to the arguments of \cite{LWY1993}, the null distribution of $\mathcal{G}$
%can be approximated  by the conditional distribution of $\mathcal{G}^*$ given the observed data,
%which can be obtained by repeatedly generating the random samples $W_i\ (i=1,\dots, n)$ while fixing the observed data.
%Thus, the critical value $\lambda_{1}(\alpha)$ can be taken as the $(1-\alpha)$-percentile of
%the conditional distributions of $\mathcal{G}^*.$

We next examine whether $\tau(v)$ varies with the mark, specifically testing the following hypothesis:
\begin{align*}
 H_{0}^C: \tau(v)\equiv\psi~\text{for}~v\in[b_L,b_U]\ \
\text{versus}\ \ H_{1}^C: \tau(v)\not\equiv \psi~ \text{for}~v\in[b_L,b_U],
\end{align*}
where $\psi$ denotes an unspecified constant.
%When $H_0^C$ is true, there is no varying trend for $\tau(v)$ with $v$,
%indicating that the effect of the treatment is constant on the time to HIV-1 infection across various IC80s.
When $H_0^C$ is true, $\tau(v)$ remains constant,
indicating that the effect of treatment on failure time does not differ significantly between different marks.

To test $H_{0}^C,$ we propose the following test statistic:
\begin{align*}
\mathcal{C}=\sup_{b_L\le v_1<v_2\le b_U} \sqrt{nh}|\widehat\tau(v_1)-\widehat\tau(v_2)|/\widehat\zeta^{1/2}(v_1,v_2),
\end{align*}
where
$$
\widehat\zeta(v_1,v_2) =nh\sum_{a\in\{0,1\}}n_a^{-2}\sum_{i:A_i=a}[\widehat\vartheta_{ai}(v_1)-\widehat\vartheta_{ai}(v_2)]^2.
$$
When $\tau(v)$ is constant over the interval $[b_L,b_U]$, $\mathcal{C}$ is close to zero.
Thus, we reject $H_{0}^C$ if $\mathcal{C}>c_{2}(\alpha)$,
where $c_{2}(\alpha)$ denotes the critical value at the significance level $\alpha$.
%
%\begin{proposition}\label{GAP2}
%Let $\mathcal{C}^*=\sup_{v\in[b_L,b_U]}B_n(v)$,
%where $B_n(v)$ denotes a centered Gaussian process
%with covariance function
%$$
%\big[h\sigma(v_1)\sigma(v_2)\big]^{-1}\text{Cov}\bigg(\frac{A}{\pi}\vartheta_{11}(v_1)-\frac{1-A}{1-\pi}\vartheta_{11}(v_1),
%\frac{A}{\pi}\vartheta_{11}(v_2)-\frac{1-A}{1-\pi}\vartheta_{11}(v_2)\bigg).
%$$
%If Conditions 1, 2 and 3' hold. then we have
%$$
%|\mathcal{C}-\mathcal{C}^*|=o_p(1).
%$$
%\end{proposition}

%To obtain $\lambda_{2}(\alpha)$, we also consider a resampling technique.
%Based on \eqref{Asym}, define
%\begin{align*}
%\mathcal{C}^*=\max_{b_L\le v_1<v_2\le b_U} \frac{h}{n}\bigg[\sum_{i=1}^n W_i\bigg\{\frac{A_i}{\widehat\pi}\Big(\widehat\vartheta_{1i}(v_1)-\widehat\vartheta_{1i}(v_2)\Big)
%-\frac{1-A_i}{1-\widehat\pi}\Big(\widehat\vartheta_{0i}(v_1)-\widehat\vartheta_{0i}(v_2)\Big)\bigg\}\bigg]^2\big/\widehat\zeta(v_1,v_2).
%\end{align*}

%Then the null distribution of  $\mathcal{C}$ can be approximated  by the conditional distribution of
%$\mathcal{C}^*$ given the observed data, and the critical value $\lambda_{2}(\alpha)$
%can be taken as the $(1-\alpha)$-percentile of the conditional distribution of $\mathcal{C}^*.$

Now we consider a Monte Carlo method to obtain the critical values $c_{1}(\alpha)$ and $c_{2}(\alpha).$
Specifically, by the proof of Theorem \ref{theo2}, we can show the following:
\begin{align}\label{Asym}
\sqrt{nh}[\widehat{\tau}_a(v)-\tau_a(v)]=\sqrt{\frac{h}{n}}\sum_{i=1}^n\frac{A_i}{\pi}\int_{0}^1
\int_{0}^{L}\frac{t-\tau_a(v)}{f_a(v)S_a(t)}K_h(u-v)N_i(dt,du)  +o_p(1)
\end{align}
uniformly in $v\in[b_L,b_U]$,
which indicates that the null distributions of $\mathcal{Z}$ and $\mathcal{C}$ can be obtained using a Gaussian approximation method.
To do that, we define
$$\vartheta_{ai}(v)=\int_{0}^1
\int_{0}^{L}[t-\tau_a(v)][f_a(v)S_a(t)]^{-1}K_h(u-v)N_i(dt,du).
$$
We modify Condition 3 as follows.

\noindent {\bf Condition 3'}.
The kernel $K(x)$ is a symmetric density function with support $[-1, 1],$
and satisfies $\mathcal{K}=\{t\mapsto K(ht+v): h>0, v\in[b_L,b_U]\}$ is a VC class with an envelope $\sup_x K(x)<\infty$.
Additionally, $h\rightarrow 0$, $nh/\log^c n\rightarrow\infty$ and $nh^5 \log n \rightarrow 0$
for some constant $c>9$.

Compared to Condition 3, Condition 3' imposes a further restriction that the class of the kernel functions should be a bounded VC class,
which is satisfied by both the uniform kernel and the Epanechnikov kernel.
We refer to \cite{GG2002} for general sufficient conditions under which $\mathcal{K}$ is a VC class.
The conditions on the bandwidth $h$ are somewhat stronger, but essentially coincide with those utilized in \cite{EM2005} and \cite{CCK2014}.
% They ensure that the rate of the approximation error vanishes faster than $\log^{-1/2}(n)$.

Based on equation \eqref{Asym},
we have the following proposition.
\begin{proposition}\label{GAP}
Define $\mathcal{G}=\sup_{v\in[b_L,b_U]} \sqrt{nh}|\widehat \tau(v)-\tau(v)|/\sigma(v)$
and $\mathcal{G}^*=\sup_{v\in[b_L,b_U]}|B_n(v)|$,
where $B_n(v)$ denotes a centered Gaussian process
with covariance function
\begin{align}\label{P3}
\big[h\sigma(v_1)\sigma(v_2)\big]^{-1}
\big[\text{Cov}\big(\vartheta_{11}(v_1),\vartheta_{11}(v_2)\big)\big/\pi
+ \text{Cov}\big(\vartheta_{11}(v_1), \vartheta_{11}(v_2)\big)\big/(1-\pi)\big].
\end{align}
If Conditions 1, 2, and 3' hold, then we have
$$
|\mathcal{G}-\mathcal{G}^*|=o_p(\log^{-1/2}n).
$$
\end{proposition}
Proposition \ref{GAP} implies that, under $H_0^C$,
\begin{align*}
\mathbb{P}(\mathcal{Z}>c_1(\alpha))\le & \mathbb{P}(\mathcal{G}^*>c_1(\alpha)-o(\log^{-1/2}n))+o(1)\\
\le & \mathbb{P}(\mathcal{G}^*>c_1(\alpha))+o(1).
\end{align*}
The second inequality follows from the anticoncentration inequality (see Lemma A.1 in \citealp{CCK2013})
and the fact that $\mathbb{E} \mathcal{G}=O(\log n)$ (\citealp{Vaart1996}, Corollary 2.2.8).
Similarly, we can show that $\mathbb{P}(\mathcal{G}>c_1(\alpha))\ge \mathbb{P}(\mathcal{G}^*>c_1(\alpha))-o(1)$.
This implies that $c_1(\alpha)$ can be estimated by simulating the distribution of $\mathcal{G}^*$.
Specifically, we repeatedly generate random samples $B_n$ with an estimated covariance function
\begin{align*}
nh\sum_{a\in\{0,1\}}n_a^{-2}\sum_{i:A_i=a}\frac{\widehat\vartheta_{ai}(v_1)}{\widehat \sigma(v_1)}\frac{\widehat\vartheta_{ai}(v_2)}{\widehat \sigma(v_2)}.
\end{align*}
Then, $c_{1}(\alpha)$ can be estimated by calculating the empirical $(1-\alpha)$-percentile of $\mathcal{G}^*$.

For $c_{2}(\alpha)$, by the continuous mapping theorem (e.g. \citealp{Vaart1996}, Theorem 1.3.6),
we see that the null distribution of $\mathcal{C}$ can be approximated by
$\mathcal{C}^*=\sup_{b_L\le v_1<v_2\le b_U}|B_n(v_1)-B(v_2)|$.
Thus, using the same arguments as above,
we can estimate $c_2(\alpha)$ by simulating the distribution of $\mathcal{C}^*$.

We conclude this section using a toy example.
For simplicity, we assume that the data are fully observed without censoring.
Given that $T(a)=T(a,V(a))$, define the total treatment effect as $\tau=\mathbb{E}[T(1)-T(0)]$.
We further define the natural direct effect (NDE) and natural indirect effect (NIE) as follows:
\begin{align*}
    \text{NDE}&=\mathbb{E}\{T(1, V(0))\}-\mathbb{E}\{T(0, V(0))\},\\
    \text{NIE}&=\mathbb{E}\{T(1, V(1))\}-\mathbb{E}\{T(1, V(0))\}.
\end{align*}
Then,  the total treatment effect can be decomposed into the natural direct effect and the natural indirect effect:
$$
\tau=\text{NDE}+\text{NIE}.
$$
Under the sequential ignorability assumption \citep{IKT2010},
and using arguments similar to those in equation \eqref{identification-tau1},
we have the following mediation formula \citep{P2001, IKT2010}:
$$
\mathbb{E}\{T(a, V(1-a))\}=\int_0^1 \mathbb{E}(T|A=a, V=v)f_{1-a}(v)dv.
$$
We consider the following example. Let $V(1)$ and $V(0)$ be drawn from a uniform distribution on the interval $[0,1]$.
Define $\tau_1(v)=3+2\sin(2\pi v)$ and $\tau_0(v)=3-2\sin(2\pi v)$.
Let $T=A\tau_1(V)+(1-A)\tau_0(V)+\epsilon$, with $\mathbb{E}(\epsilon|A, V)=0$.
Then it is not difficult to deduce that $\tau(v)=4\sin(2\pi v)\not\equiv 0$ over $[0,1]$.
However, it follows that $\tau=0$,
which indicates that using $\tau$, without controlling the mark, may not capture the treatment effect.
In addition, $\mathbb{E}\{T(a, V(1-a))\}=\mathbb{E}\{T(a, V(a))\}=3$ for $a\in\{0, 1\}$.
This implies that both NDE and NIE are equal to zero.
Therefore, it remains impossible to identify the effect with such a decomposition in this case.

\section{Simulation studies}\label{Section:SS}

In this section, we study  the finite sample performance of the proposed method.
We first independently generate $A_i$ from a binary distribution with a success probability of $2/3$.
Given $A_i$, the mark $V_i$ is generated from a beta distribution with parameters $1+A_i$ and $1+A_i$.
We set $\tau_0(v)=3-2\sin(2\pi v)$ and $\tau_1(v)=c_1+c_2\sin(2\pi v)$,
where $c_1$ and $c_2$ are specified below.
Then, the failure time $T_i$ is generated using
$T_i=A_i\tau_1(V_i)+(1-A_i)\tau_0(V_i)+\epsilon_i$,
where $\epsilon_i$ is independently generated from a truncated normal distribution with support $[-1,1]$.
Under these settings, we have $\tau(v)=(c_1-3)+(c_2+2)\sin(2\pi v)$.
The kernel function is set to be the Epanechnikov kernel function.
%The bandwidth $h$ is chosen as $h=\hat\sigma_Vm^{-1/4}$.
%which were 0.058 for $n=1000$ and and 0.053 for $n=1500$.
The censoring time is generated from an exponential distribution with mean $\mu_a$ for $A_i=a$,
where $\mu_a$ is chosen to give a censoring rate (CR) of about $40\%$ and $50\%$ with different choices of $c_1$ and $c_2$.
The results presented below are based on 5000 replications.

We first examine the consistency and asymptotical normality of the resulting estimators.
We set $c_1=3$ and $c_2\in\{-2, -1.5\}$.
We set $[b_L, b_U]=[0.2,0.8]$.
For estimation, we take the grid of 50 evenly spaced points in $[b_L, b_U]$, denoted by $\{v_1,\dots,v_k\}$.
Table~\ref{BiasRatioCP} reports the empirical biases (Bias),
the estimated standard deviations (ESD), the sample standard deviations (SSD) of $\widehat{\tau}(v)$,
and the coverage probability (CP) of the pointwise 95\% confidence band.
We observe that $\widehat\tau(v)$ is nearly unbiased,
the estimated standard deviation align well with the empirical standard deviation,
and the CP is close to the nominal level.
Figure \ref{EstimateCurve} further presents the estimated curves of $\tau(v)$,
which are close to their true curves.
Figure \ref{EstimateCurve} also presents the pointwise $95\%$ confidence bands,
which are calculated using \eqref{CI} for each point on the grid $v_k$.
We observe that the confidence bands cover the true curves.

%\begin{table}[]{@{}ccccccccccccccc@{}}
%\caption{Simulation results for Bias, ESD, SSD and CP(\%) with $c_1=3$ and $c_2\in\{-2, -1.5\}$.}\label{BiasRatioCP}
%&&\multicolumn{4}{c}{$c_2=-2$} && \multicolumn{4}{c}{$c_2=-1.5$}\\
%$v$&$n$  & Bias  & ESD & SSD & CP &&  Bias  & ESD & SSD & CP\\
%0.2	&	1000	&	0.002 	&	0.119 	&	0.123 	&	93.9 	&	&	-0.009 	&	0.120 	&	0.121 	&	94.3 	\\
%	&	1500	&	0.000 	&	0.142 	&	0.144 	&	94.5 	&	&	-0.005 	&	0.140 	&	0.143 	&	94.2 	\\
%0.4	&	1000	&	0.003 	&	0.172 	&	0.178 	&	93.1 	&	&	0.003 	&	0.167 	&	0.176 	&	93.3 	\\
%	&	1500	&	0.003 	&	0.167 	&	0.173 	&	93.5 	&	&	0.011 	&	0.163 	&	0.166 	&	93.4 	\\
%0.6	&	1000	&	0.001 	&	0.101 	&	0.103 	&	94.3 	&	&	-0.010 	&	0.102 	&	0.104 	&	94.3 	\\
%	&	1500	&	0.004 	&	0.119 	&	0.120 	&	94.5 	&	&	-0.005 	&	0.118 	&	0.120 	&	94.1 	\\
%0.8	&	1000	&	0.004 	&	0.144 	&	0.148 	&	93.8 	&	&	0.004 	&	0.141 	&	0.143 	&	94.2 	\\
%	&	1500	&	0.000 	&	0.142 	&	0.143 	&	93.9 	&	&	0.011 	&	0.139 	&	0.142 	&	93.9 	\\
%\end{table}

\begin{table}[H]\centering								
\caption{Simulation results for Bias, ESD, SSD and CP(\%) with $c_1=3$ and $c_2\in\{-2, -1.5\}$.}\label{BiasRatioCP}
\begin{tabular}{cccccccccccc}				
\hline
&&&\multicolumn{4}{c}{$c_2=-2$} && \multicolumn{4}{c}{$c_2=-1.5$}\\
\cline{4-7} \cline{9-12}
$v$&$n$  & & Bias  & ESD & SSD & CP &&  Bias  & ESD & SSD & CP\\
\hline																															
0.2	&	1000	&&	0.002 	&	0.119 	&	0.123 	&	93.9 	&	&	-0.009 	&	0.120 	&	0.121 	&	94.3 	\\
	&	1500	&&	0.000 	&	0.142 	&	0.144 	&	94.5 	&	&	-0.005 	&	0.140 	&	0.143 	&	94.2 	\\
0.4	&	1000	&&	0.003 	&	0.172 	&	0.178 	&	93.1 	&	&	0.003 	&	0.167 	&	0.176 	&	93.3 	\\
	&	1500	&&	0.003 	&	0.167 	&	0.173 	&	93.5 	&	&	0.011 	&	0.163 	&	0.166 	&	93.4 	\\
0.6	&	1000	&&	0.001 	&	0.101 	&	0.103 	&	94.3 	&	&	-0.010 	&	0.102 	&	0.104 	&	94.3 	\\
	&	1500	&&	0.004 	&	0.119 	&	0.120 	&	94.5 	&	&	-0.005 	&	0.118 	&	0.120 	&	94.1 	\\
0.8	&	1000	&&	0.004 	&	0.144 	&	0.148 	&	93.8 	&	&	0.004 	&	0.141 	&	0.143 	&	94.2 	\\
	&	1500	&&	0.000 	&	0.142 	&	0.143 	&	93.9 	&	&	0.011 	&	0.139 	&	0.142 	&	93.9 	\\
\hline		
\end{tabular} 										
\end{table}

\begin{figure}

\centering{

\includegraphics[width=6in,height=4in]{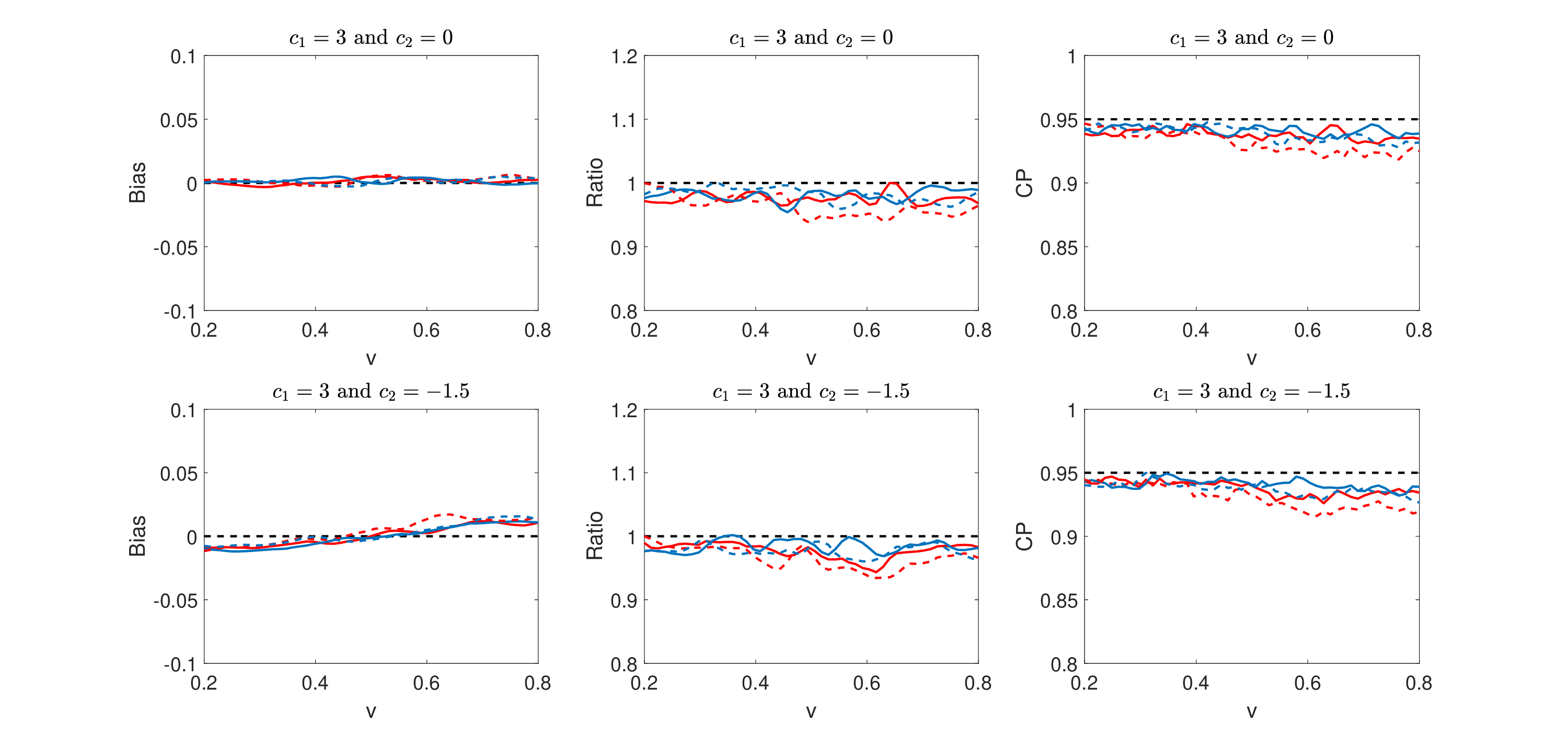}

}

\caption{\label{EstimateCurve}The simulation results for the estimating methods:
The red-solid (-dashed) lines represent the estimation results with $n=1000$ and CR$=40\%$ ($50\%$),
and blue-solid (-dashed) lines denote the estimation results with $n=1500$ and CR$=40\%$ ($50\%$).
Here, Ratio=ESD/SSD.}
\end{figure}

\begin{figure}

\centering{

\includegraphics[width=6in,height=2.5in]{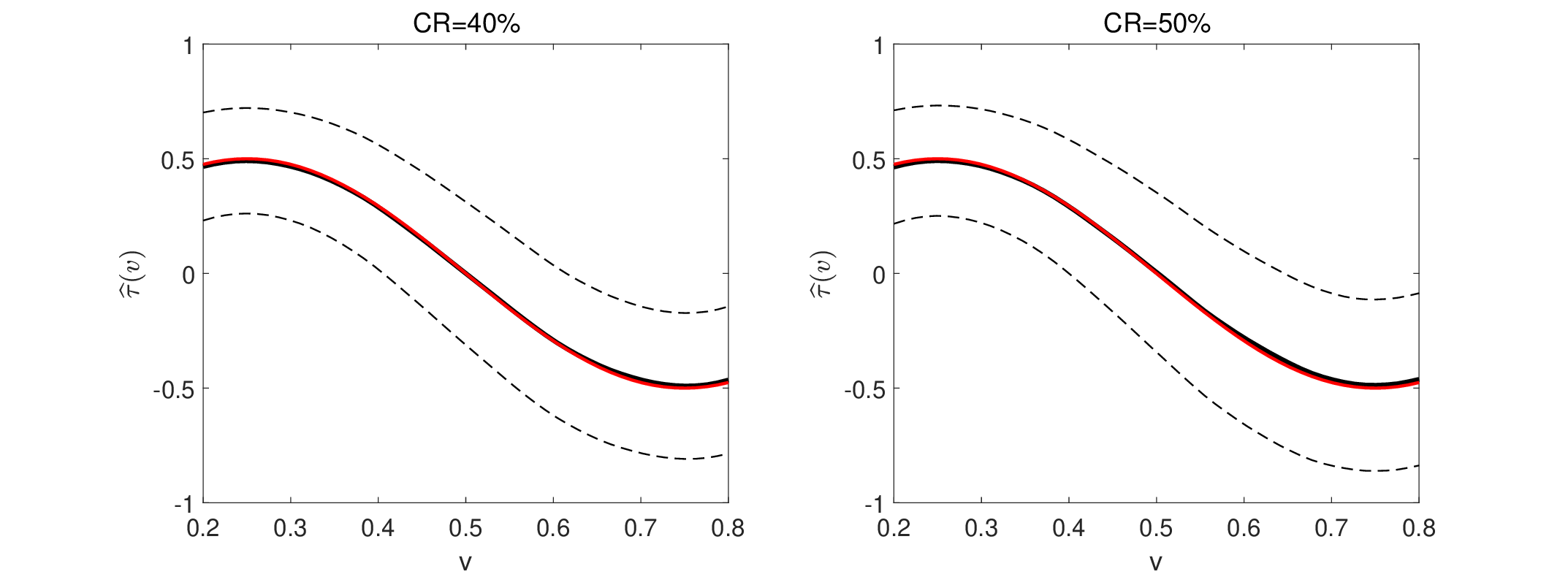}

}

\caption{\label{EstimateCurve}The estimated curves with $c_1=3$ and $c_2=-1.5$ with $n=1000$:
The black-solid lines represent the true curves,
the red-solid denote the estimated curves, and the dashed lines indicate the pointwise 95\% confidence bands.}
\end{figure}

\begin{figure}

\centering{

\includegraphics[width=6in,height=2in]{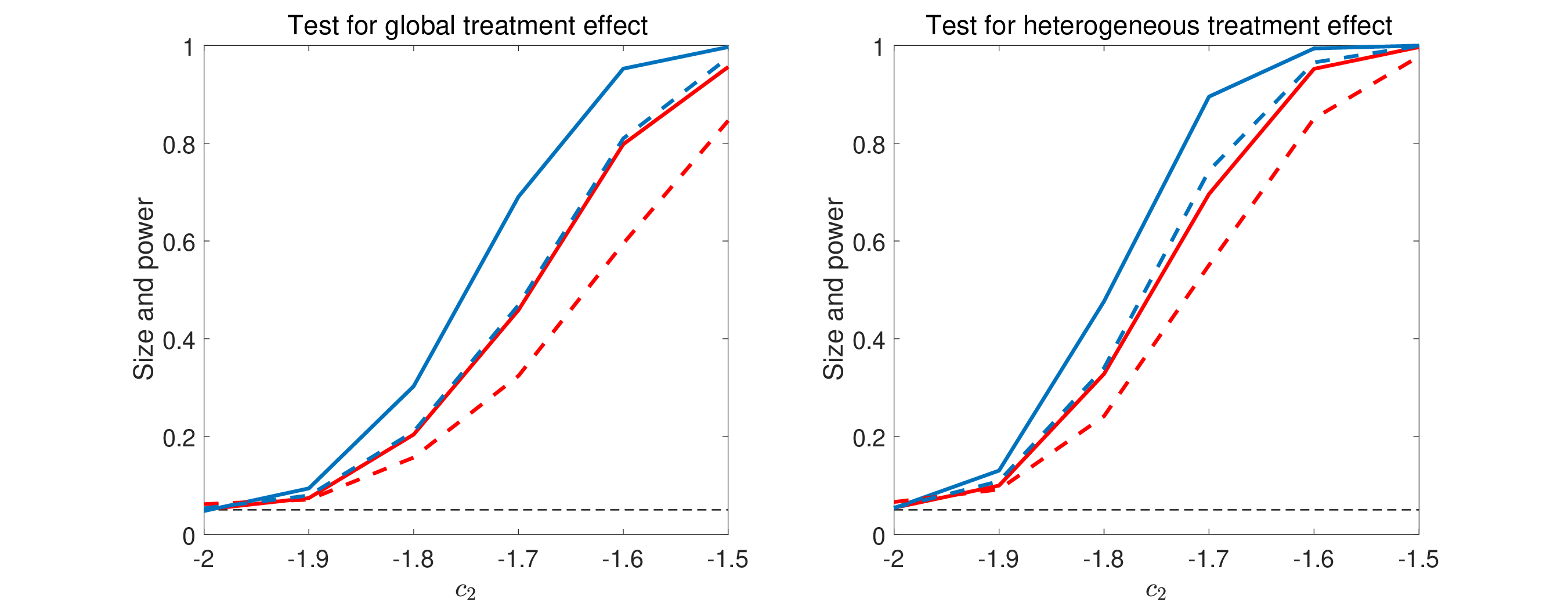}

}

\caption{\label{SizeandPower}Simulation results for the testing methods:
The red-solid (-dashed) lines represent the size and power with $n = 1000$ and CR$=40\%$ ($=50\%$),
while the blue-solid (-dashed) lines indicate the size and power with $n = 1500$ and CR$=40\%$ ($=50\%$).}
\end{figure}

Next, we evaluate the performance of the testing methods proposed in Section \ref{sec3}.
The critical values are obtained using the Monte Carlo method with 5000 simulated realizations,
and the significance level $\alpha$ is set to 0.05.
We set $c_1=3$ and $c_2\in\{-2, -1.9, \dots, -1.5\}.$
%In addition, for $H_0^G$ and $H_0^C$, we set $c_2=0$ and $c_2=2$, respectively.
When $c_1=3$ and $c_2=-2$, we have $\tau(v)\equiv 0$ on the interval $[0.2,0.8]$,
corresponding to the null hypothesis $H_{0}^G$ and $H_{0}^C$.
The case $c_2\neq -2$ is examined for the alternative hypotheses $H_{1}^G$ and $H_{0}^C$.
Figure \ref{SizeandPower} indicates that the empirical sizes of the proposed tests are close to the nominal level $5\%$,
and the powers are reasonable for detecting deviations from the null hypothesis.
We also observe that the power decreases as the censoring rate increases from $40\%$ to $50\%$;
however, it increases as $c_2$ varies from $-2$ to $-1.5$,
or when the sample size $n$ increases from $1000$ to $1500$.

\section{Real Data Analysis}\label{realdata}

We analyze data from the AMP trials: HVTN 704/HPTN 085 and HVTN 703/HPTN 081,
which were designed to determine whether bnAb can prevent the acquisition of HIV-1 \citep{Corey2021}.
The HVTN 704/HPTN 085 trial enrolled 2,687 men who were at risk of HIV infection,
while the HVTN 703/HPTN 081 trial included 1,924 women.
For each trial, subjects were randomly assigned in a 1:1:1 ratio to receive infusions of the bnAb (VRC01)
at a dose of 10 mg/kg of body weight (low-dose group), VRC01 at 30 mg/kg (high-dose group), or saline placebo,
administered at 8-week intervals for a total of 10 infusions.
We treat IC80 as the mark \citep{Corey2021,J2024}.
%The VCR01 may provide protection when the values of IC80 are low.

\begin{figure}

\centering{

\includegraphics[width=6in,height=3in]{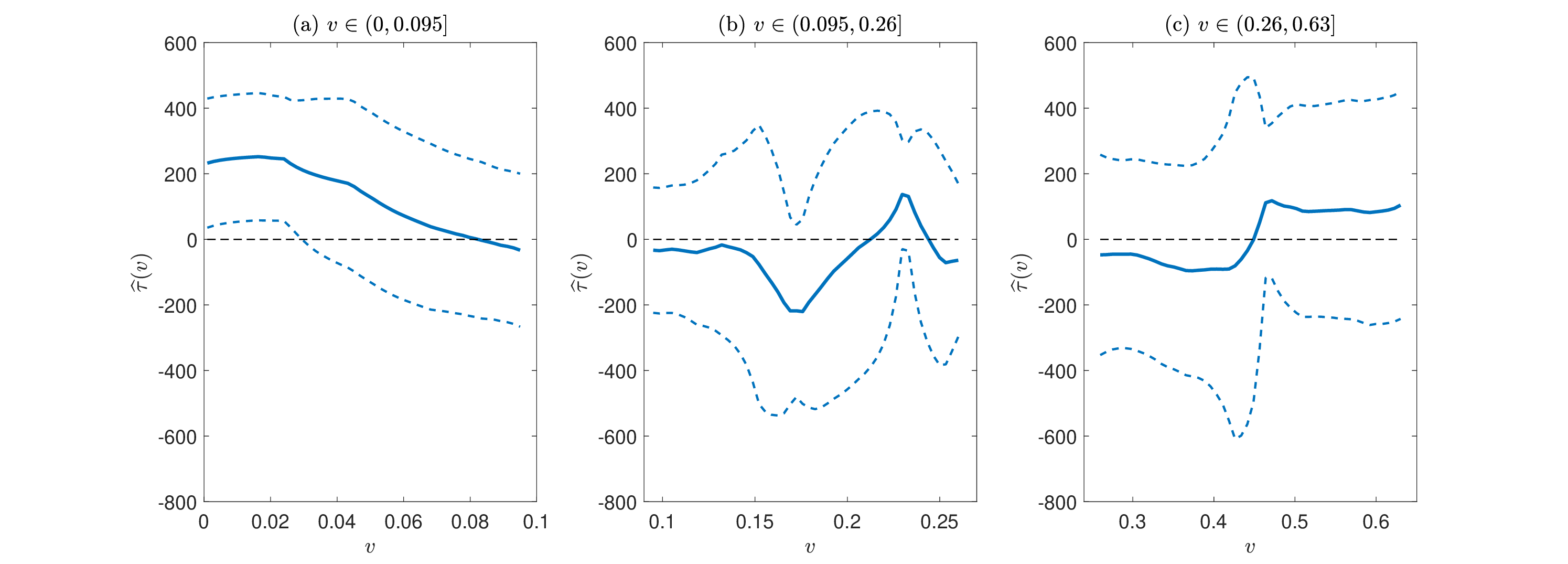}

}

\caption{\label{AMP:Est}The estimated curve for the AMP dataset:
The blue-solid lines represent the curve of $\widehat\tau(v)$,
while the blue-dashed lines indicate the pointwise $95\%$ confidence bands.}
\end{figure}

We focus on the analysis of the HVTN 704/HPTN 085 trial.
This trial identified 98 endpoints for the diagnosis of HIV infection ; however, 10 subjects had missing marks.
Our analysis is based on the 88 samples without missing marks,
which include 52 subjects in the low- and high-dose groups and 36 in the placebo group.
Each of the uncensored samples has a unique mark, ranging from $0.076$ to $39.24$.
Since the original distribution of the marks is highly right-skewed,
we scale the marks to the interval $[0, 1]$ using the function $1-e^{-x/10}$.
For analysis, \cite{Corey2021} divided the mark interval into four subintervals: $(0,1], (1,3], (3,10]$ and $(10,39.24)$,
which correspond to the scaled subintervals $(0,0.095], (0.095, 0.26], (0.26, 0.63]$ and $(0.63,1]$, respectively.
We selected $[b_L, b_U]=(0,0.095], (0.095,0.26]$ and $(0.26, 0.63]$ for our analysis.
As in simulation studies, we generate the grid of 50 evenly spaced points in $[b_L, b_U]$
and apply the Epanechnikov kernel.
The bandwidth is estimated using the method in Section \ref{BS}.
We define $A_i=1$ if the $i$th subject is in the low-dose group or the high-dose group, and $A_i=0$ otherwise.
The results are presented in Table \ref{AMP:PV} and Figure \ref{AMP:Est}.
Table \ref{AMP:PV} and Figure \ref{AMP:Est}(a) indicate that VRC01 has a significant effect on the delay in HIV-1 infection time
when the marks are within $(0,0.095]$, with a $p$-value of 0.014 for the testing $H_0^G$ versus $H_1^G$.
In addition, VRC01 shows a heterogeneous effect in different marks, with a $p$-value of 0.023.
Figure \ref{AMP:Est}(a) also indicates that the effect of VRC01 has a decreasing trend
as IC80 increases within the interval $(0,0.095]$.
This phenomenon occurs because a higher IC80 corresponds to weaker neutralizing activity,
indicating a less potent immune response induced by VRC01.
However, there is no significant efficacy of VRC01 compared to placebo
in preventing HIV-1 diagnosis, and no varying trends are observed for $v\in(0.095,0.26]$ and $(0.26,0.63]$.
These findings coincide with those reported by \cite{Corey2021}.
For comparison, we also estimate $\tau=\mathbb{E}[T(1)-T(0)]$ defined in Section \ref{sec3}.
The estimated value for $\tau$ is $-1.616$, indicating a negative effect of VCR01.
%that is different from our findings on the interval $v\in(0,0.095]$.

\begin{table}[H]\centering								
\caption{The $p$-values for the analysis of the HVTN 704/HPTN 085 trial.}\label{AMP:PV}
\begin{tabular}{ccccc }				
\hline
$v$&  & Test for $H_0^G$ && Test for $H_0^C$\\
\hline																															
$(0,0.095]$    && 0.014 && 0.023\\
$(0.095,0.26]$ && 0.508 && 0.316\\
$(0.26, 0.63]$ && 0.612 && 0.336\\
\hline		
\end{tabular} 										
\end{table}

\section{Conclusion}\label{sec-discussion}

In this article, we introduced a mark-specific treatment effect
when the failure time of interest is marked by a continuous random variable and subject to censoring.
We presented the conditions necessary for its identification
and developed a local smoothing method for its estimation.
In addition, we established a local treatment effect test, a global effect test,
and a heterogeneous treatment effect test
to comprehensively characterize the patterns of treatment effects.
To conclude the article, we discuss several interesting topics for future study.
First, our method is based on a randomness experiment;
similar ideas may be extended to observational studies.
Second, covariates are usually observed in clinical trials.
In our current work, we do not consider covariate-adjusted methods to estimate causality.
How to incorporate covariate information within our framework is another interesting topic open to discussion.

%\section*{Competing interests}
%No competing interest is declared.

\section*{Acknowledgments}

We thank Professor Yanqing Sun of University of North Carolina at Charlotte for helpful comments.
We also thank Dr. Peter Gilbert for providing data from the Antibody Mediated Prevention trials.
Lianqiang Qu's  research was partly supported by the National Natural Science Foundation of China (No. 12471256).
Liuquan Sun's research was partly supported by the National Natural Science Foundation of China (No. 12571299).

\setcounter{equation}{0}
\renewcommand{\theequation}{A.\arabic{equation}}
\section*{Appendix}
\label{sec-appenxix}
In the Appendix, we provide the proof of Theorem \ref{theo1}, Theorem \ref{theo2}, and Proposition \ref{GAP}.

{\it Proof of Theorem \ref{theo1}}.
Define
\begin{align*}
\widehat E_{a}(v)=&\frac{1}{n_a}\sum_{i: A_i=a}\int_{0}^1\int_{0}^{L}\frac{t}{\widehat S_a(t)}K_h(u-v)N_i(dt,du).
%\widehat E_{2a}(v)=&\frac{1}{n_a}\sum_{i: A_i=a}\int_{0}^1\int_{0}^{L}\frac{1}{\widehat S_a(t)}K_h(u-v)N_i(dt,du).
\end{align*}
Then $\widehat\tau_a(v)$ can be written as $\widehat\tau_a(v)=\widehat E_{a}(v)/\widehat f_{a}(v)$.
Define $G(t,u)=\mathbb{E}(N(t,u)|A=1).$
We consider the following decomposition
\begin{align}\label{wqA1}
\widehat E_{1}(v)=&\frac{1}{n}\sum_{i=1}^n\frac{A_i}{\pi}\int_{0}^1\int_{0}^{L}\frac{t}{S_1(t)}K_h(u-v)N_i(dt,du)\nonumber\\
&+\frac{1}{n}\sum_{i=1}^n\frac{A_i}{\pi}\int_{0}^1\int_{0}^{L}\bigg[\frac{t}{\widehat S_1(t)}-\frac{t}{S_1(t)}\bigg]K_h(u-v)N_i(dt,du)\nonumber\\
&+\pi\bigg(\frac{n}{n_1}-\frac{1}{\pi}\bigg)\frac{1}{n}\sum_{i=1}^n\frac{A_i}{\pi}\int_{0}^1\int_{0}^{L}\frac{t}{\widehat S_1(t)}K_h(u-v)N_i(dt,du)\nonumber\\
=& I+II+III.
\end{align}
For the first term $I$, we have
\begin{align}\label{Apen:Eq2}
I=&\int_{0}^1\int_{0}^{L}\frac{t}{S_1(t)}K_h(u-v)G(dt,du)\nonumber\\
&+\frac{1}{n}\sum_{i=1}^n\bigg(\frac{A_i}{\pi}-1\bigg)\int_{0}^1\int_{0}^{L}\frac{t}{S_1(t)}K_h(u-v)G(dt,du)\nonumber\\
&+\int_{0}^1\int_{0}^{L}\frac{t}{S_1(t)}K_h(u-v)\frac{1}{n}\sum_{i=1}^n\frac{A_i}{\pi}\big[N_i(dt,du)-G_i(dt,du)\big]\nonumber\\
=& I_1+I_2+I_3.
\end{align}
Note that $I_1=\mathbb{E}[T|A=1, V=v]f_1(v)[1+O(h^2)]$ is uniformly bounded above in $v\in[b_L,b_U]$
and $I_2$ is a sum of independent and identically distributed random variables.
An application of the central limit theory gives $I_2=O_p(n^{-1/2})$.
We next show
\begin{align}\label{Apen:Eqnew0}
\sup_{v \in [b_L,b_U]}|I_3| =O_p\bigg(\bigg(\frac{\log n}{nh}\bigg)^{1 / 2}\bigg).
\end{align}
To do that, we define
$$
W_{in}(v)=\frac{1}{n}\int_{0}^1\int_{0}^{L}\frac{t}{S_1(t)}K_h(u-v)\frac{A_i}{\pi}\big[N_i(dt,du)-G_i(dt,du)\big]
$$
and write $I_3$ as $W_n(v)=\sum_{i=1}^n W_{in}(v)$.
We divide the interval $\mathcal{S}=[b_L,b_U]$ into $M_{n}$ subintervals $I_{k}$,
with centers $v_{k}$ and length $l_{n}$.
Since $M_{n}\le 1/l_{n}$, we have $l_{n}\le 1/ M_{n}$.
Note that
\[
\begin{aligned}
\sup_{v \in \mathcal{S}}\big|W_n(v)\big|
= & \max _{1 \leq k \leq M_n} \sup_{v \in \mathcal{S} \cap I_{k}} \big|W_n(v)\big|  \\
\leq & \max _{1 \leq k \leq M_n} \sup_{v \in \mathcal{S} \cap I_{k}}\big|W_n(v)-W_n(v_k)\big|
+\max _{1 \leq k \leq M_n}\big|W_n(v_k)\big| \\
\equiv & I_{31}+I_{32}.
\end{aligned}
\]
For any $\eta>0$ , we have
\begin{align}\label{Apen:Eqnew1}
\mathbb{P}(I_{32}>\eta) \leq & \sum_{k=1}^{M_n} \mathbb{P}\big(|W_{n}(v_k)|>\eta\big)\nonumber \\
\le & \sum_{k=1}^{M_n} \bigg[\mathbb{P}\bigg(\sum_{i=1}^nW_{in}(v_k)\ge \eta\bigg)+\mathbb{P}\bigg(-\sum_{i=1}^nW_{in}(v_k)\ge \eta\bigg)\bigg]\nonumber \\
\le & \exp(-\lambda_{n} \eta)\sum_{k=1}^{M_n}\bigg[\E\exp \big(\lambda_{n} \sum_{i=1}^{n} W_{in}(v_k)\big)
+\E\exp \big(-\lambda_{n}\sum_{i=1}^{n} W_{in}(v_k)\big)\bigg]\nonumber \\
=& \exp(-\lambda_{n} \eta)\sum_{k=1}^{M_n}\bigg[\exp \big(\lambda_{n}\sum_{i=1}^n \E W_{in}(v_k)\big)
+\exp \big(-\lambda_{n}\sum_{i=1}^n\E W_{in}(v_k)\big)\bigg],
%\leq & 2 \exp(-\lambda_{n} \eta)\big[\exp \big(\lambda_{n}^{2} \sum_{i=1}^{n} \E(W_{in}(v)^{2})\big)\big] \\
%\leq & 2 \exp(-\lambda_{n} \eta)\big[\exp \big(A_{2} \lambda_{n}^{2} /(n h)\big)\big],
\end{align}
where the first two inequalities hold due to the union inequality,
and the third inequality follows from the Markov inequality.

Under Condition 1(iv) and Condition 3,
we have $|W_{in}(v)|\leq 2L\kappa_0/(s_{\min}\pi n h)$ for all $i=1, \dots, n$,
where $\kappa_0=\sup_{v}|K(v)|$.
By choosing $\lambda_{n}=(n h \log n)^{1 / 2}$,
we have
$$\lambda_{n}|W_{in}(v)| \leq 2 (\pi s_{\min})^{-1}L\kappa_0[\log n /(nh)]^{1 / 2} \leq 1 / 2$$
for a sufficiently large $n$.
Since $e^x \leq1+x+x^{2}$ for $|x| \leq1 / 2$,
we have $\exp ( \pm \lambda_{n} W_{in}(v)) \leq 1+\lambda_{n}W_{in}(v)+\lambda_{n}^{2}W_{in}(v)^2$ .
This, together with $\E(W_{in}(v))=0$ and the fact that $1+x\le e^{x}$ for $x\ge 0$, implies
\begin{align}\label{Apen:Eqnew2}
\E\big[ \exp(\pm \lambda _{n}W_{in}(v))\big] \leq 1+\lambda _{n}^{2}\E\big[ Z_{n,i}^{2}\big]
\leq \exp \big[ \E\big( \lambda _{n}^{2}Z_{n,i}^{2}\big)\big].
\end{align}
In addition, we have
\[
\E\big[W_{in}(v)^{2}\big] \leq \frac{2L^2}{n^2h^2 \pi s_{\min}} \E\bigg[K^{2}\bigg(\frac{V_{i}-v}{h}\bigg)\bigg|A_i=1\bigg] \leq C_1(n^{2} h)^{-1}[1+o(1)] .
\]
where $C_1$ denotes a positive constant.
Therefore, combining \eqref{Apen:Eqnew1} and \eqref{Apen:Eqnew2}, we obtain
\begin{align*}
\mathbb{P}(I_{32}>\eta)\le M_n \sup_{v\in \mathcal{S}} \mathbb{P}\big(|W_{n}(v)|>\eta\big)
\leq 2 M_n \exp \bigg(-\lambda_{n} \eta+\frac{C_1\lambda_{n}^{2}}{n h}\bigg).
\end{align*}
By setting
\[
\lambda_{n}=[(n h) \log n]^{1 / 2}~~ \text{ and }~~ \eta=C_2[\log n /(n h)]^{1 / 2},
\]
we obtain
\begin{align}\label{Apen:Eqnew3}
\mathbb{P}(I_{32}>\eta) \leq 2 M_n/ n^{C_2-C_1},
\end{align}
which converges to zero as $n\rightarrow\infty$ for a sufficiently large $C_{2}$
(e.g., setting $M_n\le O(n^{1/2}h^{-3/2})$ and $C_2>C_1-2$).

We now consider $I_{31}$.  Since the kernel $K(\cdot)$ is Lipschitz's continuous, we know that
\[
\begin{aligned}
\sup_{v \in \mathcal{S} \cap I_{k}}\bigg|K\bigg(\frac{V_i-v}{h}\bigg)-K\bigg(\frac{V_{i}-v_{k}}{h}\bigg)\bigg|
\leq C_{3} h^{-1} \sup_{v \in \mathcal{S} \cap I_{k}}|v-v_{k}|  \leq C_{3} h^{-1} l_{n}.
\end{aligned}
\]
Therefore, by choosing $l_{n}=h^{3/ 2} /n^{1 / 2}$, we have
\[
|I_{31}| \leq C_{3} h^{-1}l_{n}=O((n h)^{-1 / 2}),
\]
which implies that
\begin{align}\label{Apen:Eqnew4}
I_{31}=o_p((\log n/(nh))^{1/2}).
\end{align}
Combining \eqref{Apen:Eqnew3} and \eqref{Apen:Eqnew4}, we obtain \eqref{Apen:Eqnew0}.
These facts imply
$$
I=\mathbb{E}[T|A=1, V=v]f_1(v)\big[1+O(h^2)\big]+O_p\bigg(\bigg(\frac{\log n}{nh}\bigg)^{1/2}+\frac{1}{n^{1/2}}\bigg)
$$
uniformly in $v\in[b_L,b_U]$.

We next show $II=O_p(n^{-1/2})$ and $III=O_p(n^{-1/2})$.
We focus on proving $II=O_p(n^{-1/2})$, and the results for $III$ can be obtained similarly.
By Theorem 2.1 of \citet{P1991}, we have
\begin{align}\label{Apen:Eq4}
\sqrt{n}\{\widehat S_1(t)-S_1(t)\}=&
-\frac{1}{\sqrt{n}} \sum_{i=1}^n \frac{A_i}{\pi}S_1(t)\int_0^{t} \frac{dM_i^C(s)}{\bar y(s)}+o_p(1)
\end{align}
uniformly in $t\in[0,L]$.
This implies that $\widehat S_1(t)$ is a consistent estimator for $S_1(t)$ uniformly in $t\in[0,L].$
Note that
\begin{align*}
&\Bigg|\frac{1}{n}\sum_{i=1}^n\frac{A_i}{\pi}\int_{0}^1\int_{0}^{L}\bigg[\frac{t}{\widehat S_1(t)}-\frac{t}{S_1(t)}\bigg]K_h(u-v)N_i(dt,du)\Bigg|\\
%=&\Bigg|\frac{1}{n}\sum_{i=1}^n\frac{A_i}{\pi}\frac{\Delta_i Y_i[\widehat S_1(Y_i)-S_1(Y_i)]}{\widehat S_1(Y_i)S_1(Y_i)}K_h(V_i-v)\Bigg|\\
\le& \max_{t\in(0,L)}\bigg|\frac{S_1(t)}{\widehat S_1(t)}-1\bigg|\times
\frac{1}{n}\sum_{i=1}^n\frac{A_i}{\pi}\int_{0}^1\int_{0}^{L}\frac{t}{S_1(t)}K_h(u-v)N_i(dt,du).
\end{align*}
By \eqref{Apen:Eq4} and the same argument for $I$, we have $II=O_p(n^{-1/2})$
uniformly in $v\in[b_L, b_U]$.
Similarly, we can show
\begin{align}\label{Apen:Eq5}
\widehat f_{1}(v)=\int_{0}^1\int_{0}^{L}\frac{1}{S_1(t)}K_h(u-v)G(dt,du)+O_p\bigg(\bigg(\frac{\log n}{nh}\bigg)^{1/2}\bigg)
\end{align}
uniformly in $v\in[b_L, b_U]$,
which, together with \eqref{Apen:Eq2}, implies that
\begin{align*}
\widehat\tau_1(v)=&\frac{\mathbb{E}[T|A=1, V=v]f_1(v)[1+O(h^2)]}
{f_1(v)[1+O(h^2)]}+O_p\big((\log n/(nh))^{1/2}+n^{-1/2}\big)\\
=&\mathbb{E}(T|A=1, V=v)+O_p\big((\log n/(nh))^{1/2}+h^2+n^{-1/2}\big)
\end{align*}
Therefore, under the identifying condition, we have
$$
\widehat\tau_1(v)=\tau_1(v)+O_p\big((\log n/(nh))^{1/2}+h^2+n^{-1/2}\big)
$$
uniformly in $v\in[b_L,b_U]$.
% Similarly, we can show that III converges in probability to zero uniformly in $v\in[b_L, b_U]$.
Similarly, we can show
$$
\widehat \tau_0(v)=\tau_0(v)+O_p\big((\log n/(nh))^{1/2}+h^2+n^{-1/2}\big)
$$
uniformly in $v\in[b_L, b_U]$.
Therefore, we obtain
\begin{align*}
|\widehat\tau(v)-\tau(v)|\le& |\widehat\tau_1(v)-\tau_1(v)|+|\widehat\tau_0(v)-\tau_0(v)|\\
\le&O_p\big((\log n/(nh))^{1/2}+h^2+n^{-1/2}\big)
\end{align*}
uniformly in $v\in[b_L, b_U]$.
This completes the proof. 

{\it Proof of Theorem \ref{theo2}}.
To show Theorem \ref{theo2}, we use the decomposition defined in \eqref{wqA1}.
%Note that
%\begin{align*}
%\widehat\tau_1(v)=&\frac{1}{n}\sum_{i=1}^n\frac{A_i}{\pi}\int_{0}^1\int_{0}^{L}\frac{t}{S_1(t)}K_h(u-v)N_i(dt,du)\\
%&+\frac{1}{n}\sum_{i=1}^n\frac{A_i}{\pi}\int_{0}^1\int_{0}^{L}\bigg[\frac{t}{\widehat S_1(t)}-\frac{t}{S_1(t)}\bigg]K_h(u-v)N_i(dt,du)\\
%&+\pi\bigg(\frac{n}{n_1}-\frac{1}{\pi}\bigg)\frac{1}{n}\sum_{i=1}^n\frac{A_i}{\pi}\int_{0}^1\int_{0}^{L}\frac{t}{\widehat S_1(t)}K_h(u-v)N_i(dt,du).
%\end{align*}
For the second term II, using the equation \eqref{Apen:Eq4}, we have
\begin{align*}
&\frac{1}{\sqrt{n}}\sum_{i=1}^n\frac{A_i}{\pi}\int_{0}^1\int_{0}^{L}\bigg[\frac{t}{\widehat S_1(t)}-\frac{t}{S_1(t)}\bigg]K_h(u-v)N_i(dt,du)\\
=& -\frac{1}{\sqrt{n}}\sum_{i=1}^n\frac{A_i}{\pi}\int_{0}^1\int_{0}^{L}\frac{[\widehat S_1(t)-S_1(t)]}{\widehat S_1(t)S_1(t)}tK_h(u-v)N_i(dt,du)\\
=&-\frac{1}{\sqrt{n}}\sum_{i=1}^n\frac{A_i}{\pi}\int_{0}^{L}\frac{dM_i^C(s)}{\bar{y}(s)}
\frac{1}{n}\sum_{j=1}^n\int_{0}^1\int_{s}^L\frac{A_j}{\pi }\frac{t}{\widehat S_1(t)}K_h(u-v)N_j(dt,du)+o_p(1)\\
=&-\frac{1}{\sqrt{n}}\sum_{i=1}^n\frac{A_i}{\pi}\int_{0}^{L}\bigg[\int_s^Ltf_1(t,v)dt\bigg]\frac{dM_i^C(s)}{\bar{y}(s)}+o_p(1)
\end{align*}
uniformly in $v\in[b_L,b_U]$.

This implies that the second term is $O_p(n^{-1/2})$ uniformly in $v\in[b_L,b_U]$.
Similarly, we can show that the third term is also $O_p(n^{-1/2})$ uniformly in $v\in[b_L,b_U]$.
By equation \eqref{wqA1}, we have
\begin{align*}
\widehat E_{1}(v)=\frac{1}{n}\sum_{i=1}^n\frac{A_i}{\pi}\int_{0}^1\int_{0}^{L}\frac{t}{S_1(t)}K_h(u-v)N_i(dt,du)+O_p(n^{-1/2})
\end{align*}
uniformly in $v\in[b_L,b_U]$.
Additionally, we have
\begin{align}\label{Apen:Eqnew2:1}
&\widehat\tau_1(v)-\tau_1(v)\nonumber\\
=&\widehat f_1(v)^{-1}[\widehat E_1(v)-\tau_1(v)]\nonumber\\
=&\widehat f_1^{-1}(v)\bigg[\frac{1}{n}\sum_{i=1}^n\frac{A_i}{\pi}\int_{0}^1
\int_{0}^{L}\frac{t-\tau_1(v)}{S_1(t)}K_h(u-v)N_i(dt,du)+O_p(n^{-1/2})\bigg],
\end{align}
which is a sum of independent and identically distributed random variables.
Under the conditions $nh^2\rightarrow \infty$ and $nh^5\rightarrow0$,
by Slutsky's Lemma (see e.g. \citealp{Vaart1996}, Lemma 1.10.2) and equation \eqref{Apen:Eq4},
we have that for each $v\in[b_L,b_U]$,
$\sqrt{nh}\{\widehat\tau_1(v)-\tau_1(v)-\nu_0\tau_1''(v)h^2/2\}$ converges in distribution to a normal random variable with mean zero and variance $\nu_1\int_0^{\infty}(t-\tau_1(v))^2f_1(t,v)S_1^{-1}(t)dt/[f_1^2(v)\pi]$.
Similarly, we can show
\begin{align}\label{Apen:Eqnew1:2}
\widehat E_{0}(v)=&\frac{1}{n}\sum_{i=1}^n\frac{(1-A_i)}{1-\pi}\int_{0}^1\int_{0}^{L}\frac{t}{S_0(t)}K_h(u-v)N_i(dt,du)+O_p(n^{-1/2}),\\
\widehat f_{0}(v)=&\int_{0}^1\int_{0}^{L}\frac{1}{S_0(t)}K_h(u-v)G(dt,du)+O_p((\log n/(nh))^{1/2})\nonumber
\end{align}
uniformly in $v\in[b_L,b_U]$, which implies that for each $v\in[b_L,b_U]$, $\sqrt{nh}\{\widehat\tau_0(v)-\tau_0(v)-\nu_0\tau_0''(v)h^2/2\}$ converges in distribution to a normal random variable with mean zero and variance
$\nu_1\int_0^{\infty}(t-\tau_0(v))^2f_0(t,v)S_0^{-1}(t)dt/[f_0^2(v)(1-\pi)]$.
Since $\sqrt{nh}\{\widehat\tau_1(v)-\tau_1(v)\}$ and $\sqrt{nh}\{\widehat\tau_0(v)-\tau_0(v)\}$ are independent,
we have $\sqrt{nh}\{\widehat\tau(v)-\tau(v)-\nu_0\tau''(v)h^2/2\}$ converges in distribution to a normal random variable
with mean zero and variance
$\nu_1[\sigma_1^2/\pi+\sigma_0^2/(1-\pi)]$.
%$\Nu_0\{\Mathbb{E}[T(1)F_1(V|T(1))/S_1(T)]/\Pi+\Mathbb{E}[T(0)F_0(V|T(0))/S(T(0))]/(1-\Pi)\}.$
This completes the proof.

{\it Proof of Proposition \ref{GAP}}.
We apply Proposition 3.1 of \cite{CCK2014}.
For completeness, we present it as the following lemma.

\begin{lemma}
Define
$$
S_n(x, g)=\frac{1}{nh^d}\sum_{i=1}^n g(Y_i)K\big(h^{-1}(X_i-x)\big), ~~~(x, g)\in\mathcal{I}\times\mathcal{G}
$$
and
$$
\mathfrak{S}=\sup_{(x, g)\in\mathcal{I}\times\mathcal{G}} \sigma(x,g)\sqrt{nh^d}\big(S_n(x, g)-\E[S_n(x, g)]\big),
$$
where $\sigma(x,g)$ is a suitable normalizing constant,
$\mathcal{I}$ is an arbitrary Borel subset of $\mathbb{R}^d$,
and $\mathcal{G}$ denotes a class of measurable functions.

Assume that

(B1) $\mathcal{G}$ is a pointwise measurable class of functions $\mathcal{Y}\rightarrow \mathbb{R}$ uniformly bounded
by a constant $b>0$, and is VC type with envelope $b$.

(B2) $K(x)$ is a bounded and continuous kernel function on $\mathbb{R}^d$,
and such that the class of functions $\mathcal{K}=\{t\mapsto K(ht+x): h>0, x\in \mathbb{R}^d\}$
is VC type with envelope $\sup_{x}|K(x)|$.

(B3) The distribution of $X$ has a bounded Lebesgue density $p(\cdot)$ on $\mathbb{R}^d$.

(B4) $h\rightarrow0$ and $\log(1/h)=O(\log n)$ as $n\rightarrow\infty$.

(B5) $C_{\mathcal{I}\times\mathcal{G}}:=\sup_{n\ge 1} \sup_{(x,g)\in\mathcal{I}\times\mathcal{G}} |\sigma(x, g)|<\infty$.
Moreover, for every fixed $n\ge 1$ and for every $(x_m, g_m)\in\mathcal{I}\times\mathcal{G}$
with $x_m\rightarrow x\in\mathcal{I}$ and $g_m\rightarrow g\in\mathcal{G}$ pointwise,
$\sigma(x, g)\rightarrow\sigma(x, g)$.

Then for every $n\ge1$, there is a tight Gaussian random variable $B_n$ in $\ell^{\infty}(\mathcal{I}\times\mathcal{G})$
with mean zero and covariance function
$$h^{-d}\sigma(x_1, g_1)\sigma(x_2, g_2) \text{Cov}\bigg(g_1(Y_1)K\big(h^{-1}(X_1-x_1)\big),
g_2(Y_1)K\big(h^{-1}(X_1-x_2)\big)\bigg),
$$
and there is a sequence $\mathfrak{S}^*$ of random
variables such that $\mathfrak{S}^*\stackrel{d}{=}\sup_{(x,g)\in\mathcal{I}\times\mathcal{G}}B_n(x, g)$
and as $n\rightarrow \infty$,
$$
|\mathfrak{S}-\mathfrak{S}^*|=O_p\big\{(nh^d)^{-1/6}\log n +(nh^d)^{-1/4}\log^{5/4}n+(nh^d)^{-1/2}\log^{3/2} n\big\}.
$$
\end{lemma}
Using \eqref{Apen:Eqnew2:1} and \eqref{Apen:Eqnew1:2},
it suffices to verify conditions (B1)-(B5).
In this context, we define $\mathcal{Y} = [b_L, b_U]$ and $g(t) = t / S_a(t)$.
Observe that $g(t)$ is bounded above by a constant $b = L / s_{\min}$ on the interval $[0, L]$,
which ensures that (B1) holds.
Condition 3' guarantees (B2) and (B4),
while (B3) is satisfied since $V(a)$ is a continuous random variable on $[0, 1]$.
Furthermore, $\sigma^2(v)$ is continuous over the interval $[b_L, b_U]$, which verifies (B5).
By setting $g(t) =-t / S_a(t)$, we can verify conditions (B1)-(B5) in the same way.
Applying Lemma 1 leads to Proposition \ref{GAP}.
This completes the proof.

\bibliographystyle{abbrvnat}
% \bibliography{reference}

\end{document}